\documentclass[sigconf]{acmart}
\usepackage{multirow}
\usepackage{soul}
\AtBeginDocument{%
  \providecommand\BibTeX{{%
    \normalfont B\kern-0.5em{\scshape i\kern-0.25em b}\kern-0.8em\TeX}}}

\setcopyright{acmcopyright}
\copyrightyear{2022}
\acmYear{2022}
\setcopyright{acmcopyright}
\acmConference[KDD '22]{Proceedings of the 28th ACM SIGKDD Conference on Knowledge Discovery and Data Mining}{August 14--18, 2022}{Washington, DC, USA}
\acmBooktitle{Proceedings of the 28th ACM SIGKDD Conference on Knowledge Discovery and Data Mining (KDD '22), August 14--18, 2022, Washington, DC, USA} \acmPrice{15.00}
\acmDOI{10.1145/3534678.3539062}
\acmISBN{978-1-4503-9385-0/22/08}

\begin{document}

\title{Mixture of Virtual-Kernel Experts for Multi-Objective User Profile Modeling}

\author{Zhenhui Xu}
\authornote{Equal contribution. †Corresponding author.}
\email{planckxu@tencent.com}
\affiliation{%
  \institution{Tencent Inc.}
  \city{Beijing}
  \country{China}
}
\author{Meng Zhao}\authornotemark[1]
\email{jhonmenzhao@tencent.com}
\affiliation{%
  \institution{Tencent Inc.}
  \city{Beijing}
  \country{China}
}
\author{Liqun Liu}\authornotemark[2]
\email{liqunliu@tencent.com}
\affiliation{%
  \institution{Tencent Inc.}
  \city{Beijing}
  \country{China}
}
\author{Lei Xiao}
\email{shawnxiao@tencent.com}
\affiliation{%
  \institution{Tencent Inc.}
  \city{Shenzhen, Guangdong}
  \country{China}
}
\author{Xiaopeng Zhang}
\email{xpzhang@tencent.com}
\affiliation{%
  \institution{Tencent Inc.}
  \city{Shenzhen, Guangdong}
  \country{China}
}
\author{Bifeng Zhang}
\email{zbf@tencent.com}
\affiliation{%
  \institution{Tencent Inc.}
  \city{Shenzhen, Guangdong}
  \country{China}
}

\renewcommand{\shortauthors}{Zhenhui Xu et al.}
\begin{abstract}
In industrial applications like online advertising and recommendation systems, diverse and accurate user profiles can greatly help improve personalization. Deep learning is widely applied to mine expressive tags to users from their historical interactions in the system, e.g., \textit{click}, \textit{conversion} action in the advertising chain. The usual approach is to take a certain action as the objective, and introduce multiple independent Two-Tower models to predict the possibility of users' action on tags (known as CTR or CVR prediction). The predicted users' high probably attractive tags are to represent their preferences. However, the single-action models cannot learn complementarily and support effective training on data-sparse \textbf{actions}. Besides, limited by the lack of information fusion between the two towers, the model learns insufficiently to represent users' preferences on various tag \textbf{topics} well. This paper introduces a novel multi-task model called Mixture of Virtual-Kernel Experts (MVKE) to learn user preferences on various \textbf{actions} and \textbf{topics} unitedly. In MVKE, we propose a concept of Virtual-Kernel Expert, which focuses on modeling one particular facet of the user's preferences, and all of them learn coordinately. Besides, the gate-based structure used in MVKE builds an information fusion bridge between two towers, improving the model's capability and maintaining high efficiency. We apply the model in Tencent Advertising System, where both online and offline evaluations show that our method has a significant improvement compared with the existing ones and brings about an obvious lift to actual advertising revenue.
\end{abstract}




\maketitle

\section{Introduction}
User profiling is a fundamental task for many industrial applications, such as online advertising and recommendation systems \cite{yan2020learning}.
Essentially, these systems serve for good connections between users and items. 
Therefore, understanding both the users and items well is the first step. Furthermore, the behaviors, attributes, and characteristics of the users is more abundant but complicated than that of items. So an accurate and comprehensive understanding of users is the foundation of personalization. The better the user profiles we build, the more precise the connections will be.

\begin{figure*}
  \includegraphics[width=0.75\textwidth]{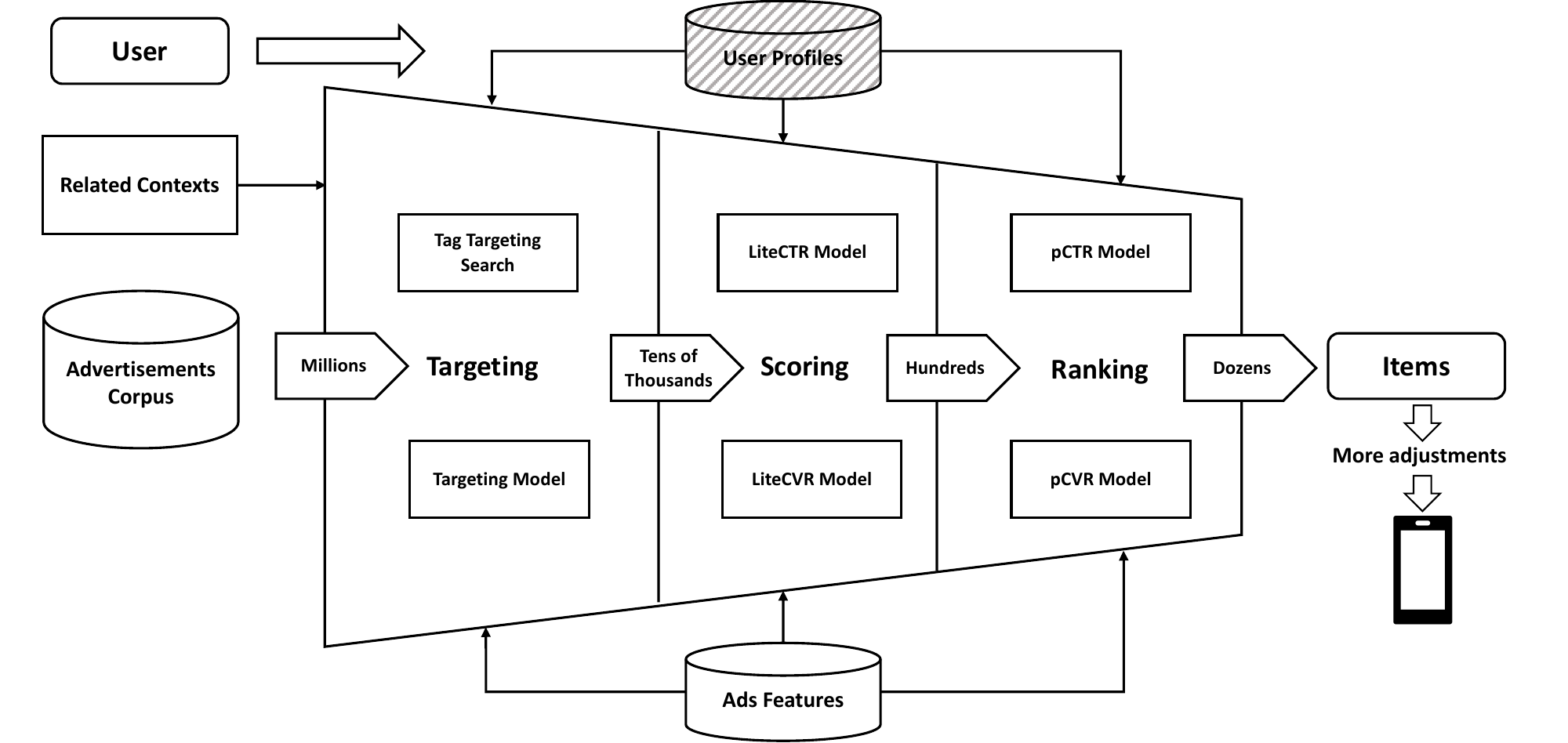}
  \caption{Tencent online advertising algorithm architecture. When one user requests ads, the system selects the best-matched item from the advertisements corpus by "\textbf{Targeting} -> \textbf{Scoring} -> \textbf{Ranking}" three main steps. User profiles serve all these main steps importantly.}
  \label{fig:ads_arch}
\end{figure*}

Similar to many famous advertising and recommendation algorithm systems in industry, e.g., Google~\cite{davidson2010youtube} and Baidu~\cite{fan2019mobius}, the architecture of Tencent advertising algorithm system is funnel-like, which is simply shown in Fig.~\ref{fig:ads_arch}. 
As is shown, when the advertising request comes, the first step is \textbf{Targeting}, that is, to collect the candidates from a massive advertising corpus (contains millions of ads) according to the user's profiles. 
The following steps are to retrieve the most appropriate ads to the user: \textbf{Scoring} and \textbf{Ranking}. 
The former is to select hundreds of ads from tens of thousands, while the latter is to select dozens out. 
In brief, both two tasks aim to select the items that the user is more likely to click and convert from respective candidates by predicting the Click-through Rate (CTR) and Conversion Rate (CVR) of user and ad.

In the advertising system, tagging is the key component of user profiling, which is labeling the users with practical, expressive tags to represent the users' preferences. It plays a vital role in all the above steps. There are many kinds of tags in user's profiles, roughly including the basic attributes (such as age, gender, education) and action-based preference tags (mined from users' inter-actions). This study focuses on mining preference tags to users, which is more challenging and crucial than the basic attributes profile building.\footnote{Unless otherwise stated, user profiling in this paper refers to tagging users.} 

The importance of tagging is mainly reflected in the following two aspects:
1) Firstly, most advertisers want their ads to reach some specified customers, e.g., \emph{Nike} intends to reach \emph{Sports} enthusiasts, and \emph{BMW} wants to get \emph{Car} enthusiasts. 
Thus, when placing the ads to the corpus, the advertisers always bind them with relevant tags. 
When one user sends an ad request, the ads bound with the same tags as this user are searched fast by \textbf{Targeting} module, as the candidates of the next steps.
Targeting by tags is the most efficient way to search the candidates, and it is also the most crucial downstream application of tagging.
2) Secondly, tagging is an essential part of feature engineering in the following \textbf{Scoring} and \textbf{Ranking} steps. 
As usual, the models in these steps are mainly to predict CTR or CVR of the user towards candidate ads, where the user's tags serve as very important features. 
The more accurate and diverse tags involved, the more precise pCXR (predicted Click-Through or Conversion Rate) will be.
In general, tagging based on users' historical actions to find more potential preference tags is much beneficial to all \textbf{Targeting}, \textbf{Scoring} and \textbf{Ranking} modules. 

However, there are three main challenges while tagging in advertising systems:
\begin{itemize}
\item \textbf{C1}: 
Tencent serves more than billions of active users per day on WeChat, QQ, and other platforms. 
Tagging such billions of users with thousands or even millions of meaning-defined tags is much challenging.

\item \textbf{C2}: The user's preferences are \textbf{topic-related}. For example, one user may prefer not only \emph{Sports}-related ads but also \emph{Car}-related ones. So it is challenging to represent the user's multiple-preference on various topics accurately and effectively.

\item \textbf{C3}:  The user's preferences are \textbf{action-based}. In advertising, \emph{click} and \emph{conversion} are two main types. 
When displaying an ads, actions are usually in sequential pattern: \emph{"impression -> click -> conversion"}. The correlation and difference of these actions must be considered seriously while mining corresponding preferences.
\end{itemize}

\begin{figure}[!htbp]
  \includegraphics[width=0.42\textwidth]{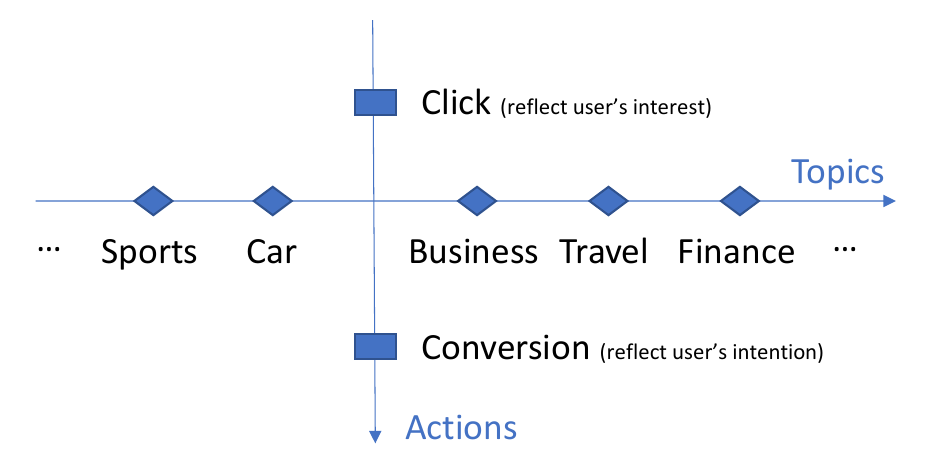}
  \caption{Two-dimensional modeling objectives. The users' preferences are multi-faceted on different topics and actions. For example, a user always \emph{clicks} \emph{Car} ads, and also always buys (one kind of \emph{conversion}) \emph{Sports} goods from ads.}
  \label{fig:two-dim}
\end{figure}

Intuitively, some existing methods can solve part of the challenges. 
As for \textbf{C1}, Two-Tower models~\cite{huang2013learning,yi2019sampling} can efficiently predict the matched tags for large-scale users, where one tower is to learn user representation, and the other tower is to learn tag representation. 
When predicting, the two towers individually output all user embeddings and all tag embeddings respectively, then the dot-products of embedding pairs are calculated as ranking scores.
However, the two towers are relatively separate, resulting in insufficient feature interaction, affecting the modeling effect. 
Therefore, facing \textbf{C2}, it is hard to represent users' multi-preferences on various topics, for the user tower produces only one fixed embedding without being aware of different topic information contained in the tag tower.
To solve \textbf{C3}, multi-task learning models, such as MMoE~\cite{ma2018modeling}, ESMM~\cite{ma2018entire}, are widely used in industrial unified modeling tasks. Nevertheless, without specialized thinking and structure design, these models are also not enough to address users' multiple topic-related preference problems, as discussed in \textbf{C2}.
In sum, under the efficiency requirement in \textbf{C1}, there are two dimensional orthogonal objectives for user profile modeling not being well solved: \textbf{topic-related} preference in \textbf{C2} and \textbf{action-based} preference in \textbf{C3}, as shown in Fig.~\ref{fig:two-dim}.

To address the challenges mentioned above, we propose a novel multi-task model named Mixture of Virtual-Kernel Experts (MVKE), which can build users' profiles efficiently, accurately, and comprehensively. 
It is also a Two-Tower model in nature. While labeling tags to billions of users with low time consumption in prediction, we design a new multi-task learning structure to handle multiple modeling objectives. 
There are two critical components in MVKE: Virtual-Kernel Expert (VKE) layers and Virtual-Kernel Gate (VKG) layers. 
The VKE aims to represent the different facets of the users' preference, by defining multiple specialized layers and various combinations of user feature inputs. In each VKE, there is a topic-related and learnable virtual kernel to guide user features interaction. 
The VKG is designed to combine the VKE outputs selectively according to the tag embedding and virtual kernels. In this way, one user would have specific preference representations for different tags. 
Overall, the model implicitly assigns users' preferences learning into different VKEs for two-dimensional modeling objectives. Each VKE can capture different facets of preferences, and VKG refines the final representation varying on different topics and actions.
Therefore, a Mixture of Virtual-Kernel Experts helps multi-objective user profile modeling with good effectiveness and efficiency.

The contributions of this paper are as follows:
\begin{itemize}
    \item We propose a Mixture of Virtual-Kernel Experts (MVKE) model, which can mine diverse and accurate tags for a large number of users by specialized structures on multi-objective modeling.
    \item We introduce a new concept of VKE in this model to learn different facets of user's preference, and VKG to refine the final representation. 
    \item MVKE meets the needs of multi-objective learning and applies a new structure to enhance the information fusion between two towers, also maintaining prediction efficiency.
    \item Our Method has brought huge real benefits for the improvement of online advertising performance in Tencent Ads.
\end{itemize}

\section{Related Works}
Our work is highly related to two fields: online prediction models and multi-task learning models. On the one hand, our approach is to handle the tag prediction task and improved from the Two-Tower framework, both related to the online prediction field. On the other hand, MVKE is also under the multi-task learning framework, whose representative works are introduced below.

\subsection{Online Prediction Models}
Online prediction tasks \cite{ke2019deepgbm} represent a certain type of tasks playing the essential role in many real-world industrial applications, such as click prediction \cite{graepel2010web, zhang2016deep, qu2016product, guo2017deepfm, lian2018xdeepfm} in sponsored search or advertising, content ranking \cite{agichtein2006improving, cao2007learning, burges2010ranknet} in web search, content optimization~\cite{wang2015collaborative, cheng2016wide, covington2016deep} and user profile modeling~\cite{kanoje2015user,lashkari2019survey,yan2020learning,gu2020hierarchical} in recommender systems or online advertising, travel time estimation~\cite{wang2018learning, li2018multi} in transportation planning, etc. 

Applying neural networks (NN) for online prediction tasks helps reach a better performance than traditional Gradient Boosting Decision Trees (GBDT) based approaches. 
Many recent works have employed NN in prediction tasks, and they mainly focused on the sparse categorical features for widely-exists items in industry. 
The evolution of these NN models is mainly focusing on automatic feature interaction, from Wide\&Deep~\cite{cheng2016wide}, DeepFM~\cite{guo2017deepfm} to xDeepFM~\cite{lian2018xdeepfm} models. Recently, \cite{he2014practical,zhu2017deep,ling2017model,ke2019deepgbm} propose models to combine GBDT and NN together, and handle tabular input space better. These models are the base for online prediction tasks, mainly working for feature extraction, interaction, and reasoning. However, they are not highly related to real business scenarios, only employed as the basic structure of industry models. 

In industry, for better effectiveness or higher efficiency, the research field on business-related model design is more and more active.
On the one hand, the Two-Tower model is proposed and widely used, benefiting from its high prediction efficiency.
Most early, Microsoft~\cite{huang2013learning} proposes a semantic DSMM model, which is regarded as the first Two-Tower model and gets good performance in the web search task. 
After it, not limited in the semantic modeling field, the Two-Tower model is applied to solve many other types of tasks \cite{yan2020learning,yi2019sampling,huang2020embedding} well, such as recommendation, advertising, because of its good efficiencies.
Recently, Google~\cite{yi2019sampling} enriches the Two-Tower model framework via a study on sampling-bias-corrected, and leads to recommendation quality improvement for YouTube.
And Tencent~\cite{yan2020learning} matches the tags to the user by Two-Tower model and shares many practical experiences.
On the other hand, to more effectively utilize the user's preference in the model, many works~\cite{zhou2018deep,cen2020controllable,Zheng2010User,pi2020search} on model structure improvement are proposed. For example, Alibaba proposes end-to-end Deep Interest Network (DIN)~\cite{zhou2018deep}, SIM~\cite{pi2020search} and ComiRec~\cite{cen2020controllable} to better capture the user interests on CTR predictions. These models also give good generalization insights for user preferences applications in industry.


\subsection{Multi-task Learning Models}
For multiple business-related tasks need, there raises a research climax for multi-task learning (MTL) models in the industry. 
Most early, hard parameter sharing~\cite{caruana1997multitask} gives a straightforward way to construct an MTL model, most of whose parameters for each task are shared, and only top layers are different, thus may make conflicts among the tasks. 
Lately, cross-stitch~\cite{misra2016cross} and sluice network~\cite{ruder2017sluice} both propose to learn weights of linear combinations to fuse representations from different tasks selectively.
Recently, there have been some studies on employing gate networks and attention mechanisms to fuse task-specific information well. Google proposes the MMoE model~\cite{ma2018modeling} based on MoE~\cite{jacobs1991adaptive} to obtain good performance with good use of task-customized gates network. Tencent presents the PLE model~\cite{tang2020progressive} with the adoption of a progressive routing mechanism and achieves a consistent improvement on all the tasks.

Besides, some works focus on the design of task goals. 
Alibaba~\cite{ma2018entire} proposes the ESMM model with a new-designed pCTCVR goal in the advertising field, to handle the sample selection bias and improve conversion rate prediction. 
Besides, some other works, such as DUPN~\cite{ni2018perceive} and ESM2~\cite{wen2019conversion} model, also applied MTL in the recommendation for some specific business goals. DUPN is proposed to build better general user representations, and ESM2 focuses on modeling more detailed user actions. 
More recently, Google~\cite{zhao2019recommending} applied MTL to video recommendations on Youtube, supplying many practical experiences. 
These models are designed deeply for real application scenarios and less focused on MTL itself, so their MTL usage is relatively straightforward and simple.


\section{Preliminaries}
In this section, we first formalize our real business problem with some definitions. Then, a basic solution is introduced, which is running in the online system.

\subsection{Formalization}
We first introduce some essential definitions and then formalize our problem to investigate.

Assuming there is a set $\mathcal{U}$ of users, in which each user $u \in \mathcal{U}$ is labeled with a set $\mathcal{T}_u(u)$ of its preference tags. Meanwhile, there is also a set $\mathcal{A}$ of advertisement items, in which each advertisement item $a \in \mathcal{A}$ is labeled with a set $\mathcal{T}_a(a)$ of its attribution tags. Besides, for each user $u$, there are a click action set $\mathcal{C}(u)$ and a conversion action set $\mathcal{V}(u)$, in which the elements are advertisement items. 

In online advertising, there are two important tasks: advertisement understanding and user profiling. 
Apparently, advertisement understanding is to build the set $\mathcal{T}_a(a)$, and user profiling is to build the set $\mathcal{T}_u(u)$. 
In this work, we aim to build user profiles, and more specifically, tagging for users. Here we define some essential concepts:

\noindent\textbf{User Interest Tagging.} If one user always clicks a category of ads, we call the user is interested in this category, labeled by a tag. Therefore, we define tagging from the click actions as User Interest Tagging. Formally, given the user click set $\mathcal{C}(u)$ and the advertisement tag set $\mathcal{T}_a(a)$, User Interest Tagging is to build an interest tag set $\mathcal{T}_u^C(u)$ for users.

\noindent\textbf{User Intention Tagging.} Similarly, if one user always converts a category of ads, e.g., one user always buys \emph{Sports}-related items from the ads, we call the user has an intention to \emph{Sports}, and the tag \emph{Sports} is labeled to the user. Therefore, we define tagging from the conversion actions as User Intention Tagging. Formally, given the user conversion set $\mathcal{V}(u)$ and the advertisement tag set $\mathcal{T}_a(a)$, User Intention Tagging is to build an intention tag set $\mathcal{T}_u^V(u)$.

\noindent\textbf{User Tagging.} In sum, User Tagging aims to mine the users' preferences from their historical actions and record them in the form of tags. Formally, given the user action sets: $\mathcal{C}(u)$, $\mathcal{V}(u)$, and the advertisement tag set $\mathcal{T}_a(a)$, User Tagging is to build user tag set $\mathcal{T}_u(u)$ for each user $u \in \mathcal{U}$. Note that more similar tagging based on other actions is also supported, without unnecessary details here.

    
    

\subsection{Basic Solution}
\label{sec:basic-solution}
In practice, it is used to build two sets, $\mathcal{T}_u^V(u)$ and $\mathcal{T}_u^C(u)$ individually, and then merge them to the user profiles $\mathcal{T}_u(u)$ together. Basically, there are two models: the interest model and the intention model. The former is to label interest tags for users and the latter is to label intention ones.

\begin{figure}[!htbp]
  \includegraphics[width=0.45\textwidth]{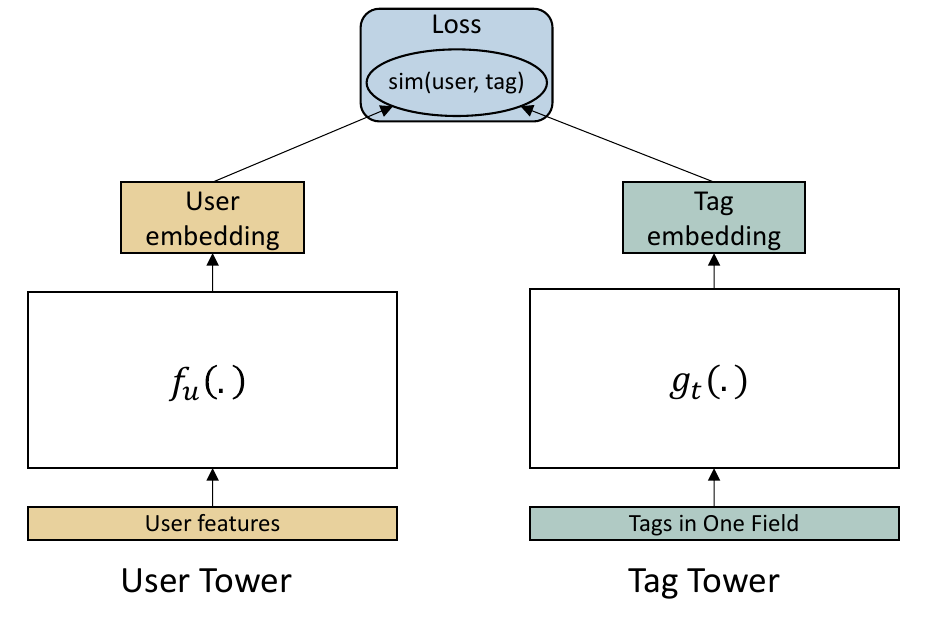}
  \caption{Basic Two-Tower model architecture on single tagging task.}
  \label{fig:basic-model}
\end{figure}

\begin{figure*}
  \includegraphics[width=0.78\textwidth]{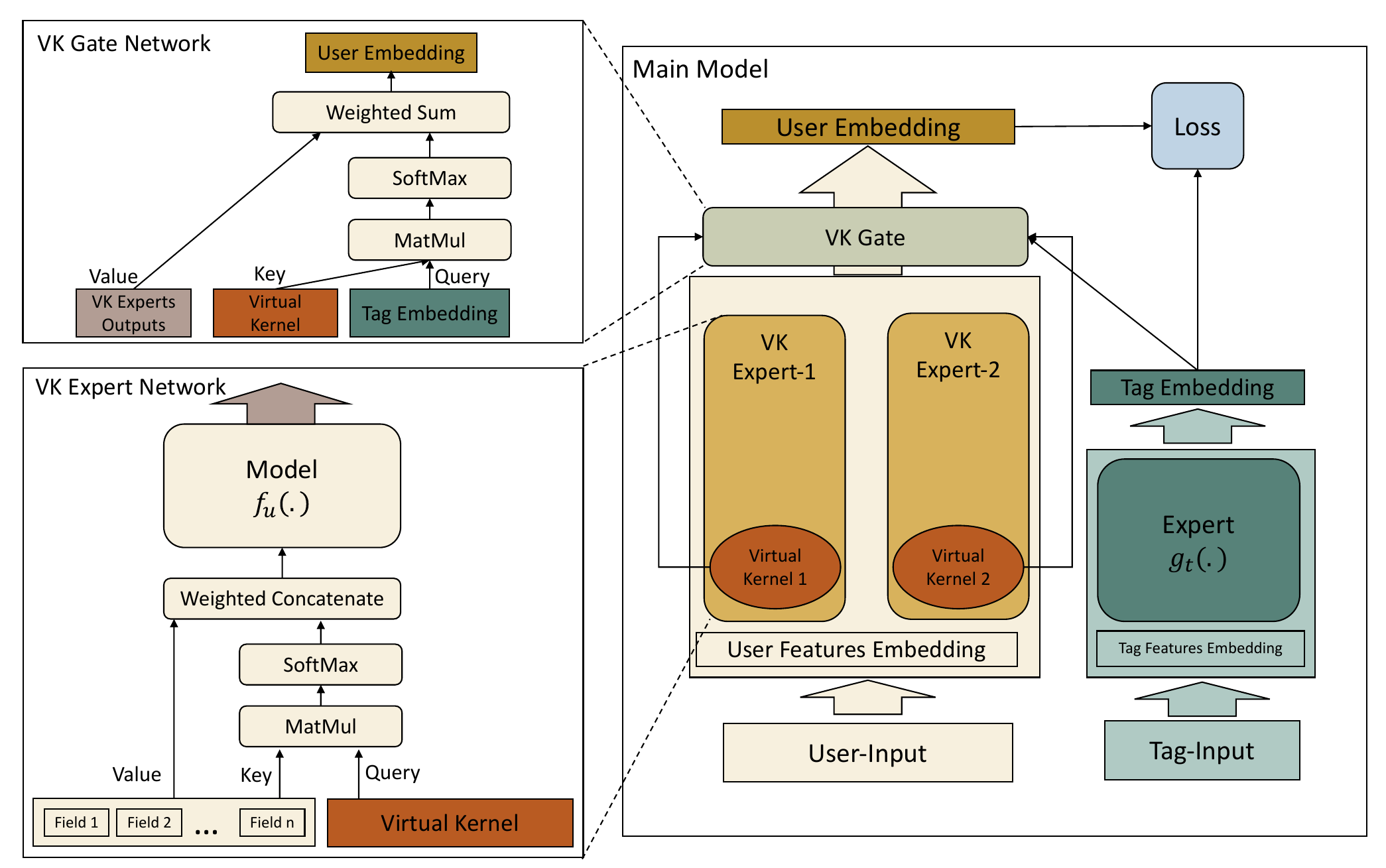}
  \caption{MVKE model architecture on single tagging task. The right part is the primary architecture of the model, where set two VKEs. The left part shows the details of VKE and VKG structures.} 
  \label{fig:mvke-single}
\end{figure*}

Similar to the prediction models in advertising, the interest model is to predict the Click-through Rate of \emph{<user, tag>} pair, and the intention one is to predict its Conversion Rate. 
As shown in Fig.~\ref{fig:basic-model}, the model contains two main parts, the user tower $f_u(\cdot)$ and tag tower $g_t(\cdot)$. 
Given a click or conversion pair <$u_i, a_i$> for training, the user tower inputs the user's various features, such as basic attributes (e.g., age, gender), and historical behavior features(e.g., reading, shopping). 
Borrowed from online prediction models like DeepFM~\cite{guo2017deepfm}, there are also multiple feature fields designed for different kinds of feature input in the model.
Besides, the tag tower inputs with the ad tag set $\mathcal{T}_a(a_i)$, and all the tags are fed into one feature field.
Finally, each tower is used to generate a vector: user embedding $E_u$ and tags embedding $E_T$ for each. Formally,
\begin{equation}\small
\begin{aligned}
    E_{u_i} = f_u(u_i^1, u_i^2, ..., u_i^m; \Theta_u), \\
    E_{T_i} = E_{a_i} = g_t(\mathcal{T}_a(a_i); \Theta_t).
\end{aligned}
\end{equation}
And the model estimation score is computed by the cosine similarity of two embeddings, and the loss function $ \mathcal{L} $ is binary cross-entropy loss supervised by the label $y_i$, which depends on whether the user $u_i$ clicks or converts the ad item $a_i$. In form,
\begin{equation}\small
\begin{aligned}
    p_i &= \sigma(cos(E_{u_i}, E_{T_i})), \\
    \mathcal{L} &= \mathcal{L}_{BCE}(\boldsymbol{y}, f_u(\boldsymbol{u}; \Theta_u) \cdot g_t(\mathcal{T}; \Theta_t)) \\ &= \sum_{i}(y_ilog(p_i) + (1-y_i)log(1-p_i)),
\end{aligned}
\end{equation}
where $\sigma$ is the $sigmoid$ activation function, $\boldsymbol{y}$ is the labels, $\boldsymbol{u}$ is the user features inputs, $\mathcal{T}$ is the tags input, and $\Theta$ is the model parameters.
When predicting the click rate of <$u_i, t_j$>, the user tower has no change, but the tag tower is input with only one single tag $t_j$, denoted as
\begin{equation}\small
    E_{t_j} = g_t(t_j; \Theta_t).
\end{equation}


The above solution is widely used but cannot produce multi-objective tags effectively. It needs to build two independent models to mine interest and intention tags separately, and this vanilla Two-Tower model sacrifices a certain degree of accuracy for efficiency. In the next section, we propose one unified model called MVKE to achieve the goal better, unitedly mine diverse and accurate tags, save computing resources, and speed up efficiency.

\section{MVKE Model}
To address aforementioned challenges, we propose a novel model named Mixture of Virtual-Kernel Experts (MVKE), which performs well in many scenarios.
\subsection{on Single Task}
In order to show its structure clearly, we first introduce MVKE on the single task goal. The architecture of MVKE is shown in Fig.~\ref{fig:mvke-single}, which is still like a Two-Tower model, but there is a "bridge" between the towers. 
We can find that the inside structure of the tag tower is nearly the same as the one in the basic solution, but the user tower is much different.

Basically, two key components help MVKE break the barrier between two towers: Virtual-Kernel Experts (VKE) and Virtual-Kernel Gate (VKG). One VKE only focuses on one facet of the users' preferences with the help of the corresponding virtual kernel. And the VKG is an attention-based weighted gate, which is to combine the VKE outputs selectively to the final user representation. In other words, the modeling of VKE is differentiated and diverse, and VKG combines VKE outputs according to different tags' attention on virtual kernels.

There are multiple VKEs in the user tower, but all the VKEs share the same feature input layer. The number of them is a hyper-parameter that can be set flexibly. 
Taking the k-th VKE as an example, there is a Virtual Kernel inside, denoted as $W_{VK}^k$. 
The virtual kernel is a learnable variable, and input as "Query" in attention mechanism of VKE. 
The "Key" and "Value" here are both user features embedding $E_{uf_i}$. 
In form, the first step in VKE is a non-linear transformation for each input respectively, which is defined as
\begin{equation}\small
\begin{aligned}
    Q &= \sigma(W_Q^T W_{VK}^k + \boldsymbol{b}_Q)~, \\
    K &= \sigma(W_K^T E_{uf_i} +\boldsymbol{b}_K)~, \\
    V &= \sigma(W_V^T E_{uf_i} +\boldsymbol{b}_V)~,
\end{aligned}
\end{equation}
where $\sigma(\cdot)$ denotes applying activation function, $W$ and $\boldsymbol{b}$ mean the weights and biases in linear layer. Then it computes the attention weight to guide the combination by input features:
\begin{equation}\small
    C_{VKE}^k = \text{softmax}(\frac{Q K^T}{\sqrt{d_{K}}}) * V.
\end{equation} 

In the top layers, a simple method of aggregating is a weighted sum of each feature field embedding. Benefiting from the strong flexibility of VKE, we can also choose to superimpose some complex interaction structures, such as DeepFM, xDeepFM, on weighted user field embedding, denoted as
\begin{equation}\small
     E_{u_i}^k = f_u^k(C_{VKE}^k).
\end{equation} 

Each VKE outputs a compact embedding $E_{u_i}^k$ under the guidance of virtual kernel, but it is implicit and has no practical meaning. An attention-based soft gate network, called VKG, is designed to generate a tag-specific user representation. VKG is also based on attention mechanism, inputs with all virtual kernels $W_{VK}$ ("Key"), tag embedding $E_{T_i}$ ("Query") and all VKE outputs $E_{u_i}$ ("Value"). Formally, the attention weight is computed by non-linear transformed $Q(E_{T_i})$ and $K(W_{VK}^k)$, and then output the weighted sum of $V(E_{u_i}^k)$, as the final user embedding $E_{u_i}$, which is defined as
\begin{equation}\small
    E_{u_i} = \sum_{k} \text{softmax}(\frac{Q K^T}{\sqrt{d_{K}}}) * V.
\end{equation} 
The loss function is also similar to the one defined in the basic solution, computed by user embedding $E_{u_i}$ and tag embedding $E_{T}$.

Looking back on the overall structure, we can easily find that the virtual kernel plays an important role in MVKE. It is used both in VKE and VKG, thus can be regarded as a bridge linking user and tag. VKEs focus on user's implicit preferences, which are aggregated in VKG into user representation corresponding to tags (maybe as aforementioned "Sports") with practical meaning. In some sense, virtual kernel determines one specific learning space of the user's preferences. Moreover, the whole of virtual kernels can be combined and mapped to different real tag spaces. With the help of virtual kernel acting as a bridge, the user and tag features can interact with each other better.

\subsection{on Multiple Tasks}

When expanding from one single task to multiple ones, we can set much more VKEs in the model, and different subsets of VKEs would serve different tasks. It's worth noting that one expert can serve for one or more tasks. The overview of MVKE on multiple tasks is shown in Fig~\ref{fig:mvke-multi}, including two tasks.
One is to predict if the user clicks the tags (interest tagging), and the other is to predict if the user converts (intention tagging).
As one VKG serves one specific task, the number of VKGs is the same as the number of tasks. In principle, we can assign any number of VKEs for a task freely. However, considering the real application needs, there are some suggestions for setting:

\begin{figure}
  \includegraphics[width=0.44\textwidth]{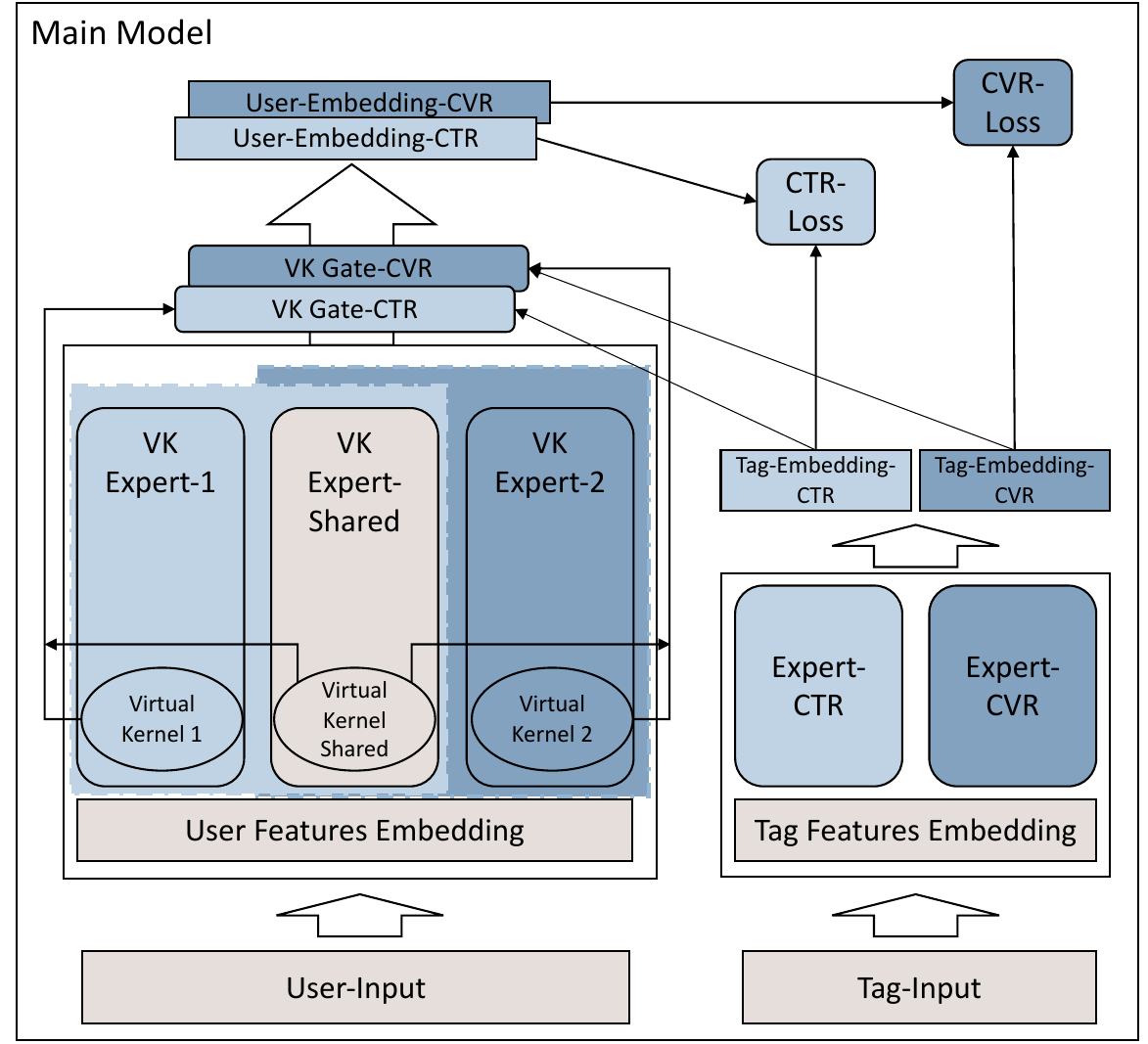}
 \vspace{-10pt}
  \caption{MVKE model architecture on multiple tagging tasks: Interest Tagging and Intention Tagging.}
  \label{fig:mvke-multi}
\end{figure}
1) Set both independent VKEs and shared VKEs for these tasks, to guarantee the specialization and generalization of the model on different tasks. Specifically, the output of each task is determined by multiple VKEs, and a shared VKE can affect the results of multiple tasks simultaneously in consideration of complementary between tasks. Meanwhile, in order to ensure the difference of training for different tasks, at least one independent VKE that only serves one task is also necessary. 

2) For tasks in one exampled sequential pattern: \emph{"impression -> click -> conversion"}, the deeper task can contain most of VKEs serves for the shallow tasks, to utilize more base task information.
In our real business scenarios, MVKE is applied for both interest tagging (pCTR-like goal) task and intention tagging (pCVR-like goal) task. As \emph{"click"} is the previous action of \emph{"conversion"}, we totally set 5 VKEs, where 1st to 3rd ones serve for interest tagging, and 2nd to 5th ones serve for intention tagging.

The final loss function defined on multiple tasks is the sum of individual task losses, which is defined as:
\begin{equation}\small
    \mathcal{L}_{MTL} = \mathcal{L}_{ctr} +\mathcal{L}_{cvr}. 
\end{equation}
More specifically, 
\begin{equation}\small
\begin{aligned}
    \mathcal{L}_{ctr} =&\mathcal{L}(\boldsymbol{y}_{ctr}, m(\{f_u^k\}_{k \in K_{s} \cup K_{ctr}}) \cdot g_t^{ctr}), \\ 
    =&\mathcal{L}(\boldsymbol{y}_{ctr}, m(\{f_u^k(\boldsymbol{u}; \Theta^k_u)\}_{k \in K_{s} \cup K_{ctr}}) \cdot g_t^{ctr}(\mathcal{T}; \Theta^{ctr}_t)), \\
    \mathcal{L}_{cvr} =&\mathcal{L}(\boldsymbol{y}_{cvr}, m(\{f_u^k\}_{k \in K_{s} \cup K_{cvr}}) \cdot g_t^{cvr}), \\ 
    =&\mathcal{L}(\boldsymbol{y}_{cvr}, m(\{f_u^k(\boldsymbol{u}; \Theta^k_u)\}_{k \in K_{s} \cup K_{cvr}}) \cdot g_t^{cvr}(\mathcal{T}; \Theta^{cvr}_t)), \\
\end{aligned}
\end{equation} 
where $m(\cdot)$ is applying combination projection for multiple VKEs, $\Theta$ is the model parameters, $K_s$ is a set in which the VKEs are shared for both two tasks, and $K_{ctr}$ or $K_{cvr}$ is the set of task-specific VKEs.

In fact, MVKE has good flexibility and a wide range of applications, so tagging from users' more various actions is also supported. For example, there are multiple types of conversions defined in oCPA (optimized cost-per-action) advertising, such as \textit{Subscribe, Activate, Purchase}, etc. If we expand aforementioned general sequential action patterns into more detailed longer ones like \textit{"impression -> click -> subscribe -> purchase"}, we can still employ MVKE to handle these tasks. The usage of MVKE on these tasks is the same, which is not elaborated here due to limited space.

\begin{table}[!htbp]
\caption{Inference cost comparison among Single Tower, Two-Tower and MVKE model.}
\label{tab-complexity}
\vspace{-12pt}
\centering
\small
\setlength\tabcolsep{3.5pt}
\begin{tabular}{c|ccc}\toprule
\textbf{Method} & \textbf{Time Compl.} & \textbf{Space Compl.}  & \textbf{Actual Time} \\ \hline
Single Tower & $O(|\mathcal{U}|*|\mathcal{T}|)$ & $O(1)$ & 60+ days\\
Two-Tower & $O(|\mathcal{U}|)$ & $O(|\mathcal{U}|)$ & 3-4 hours\\ 
MVKE & $O(|\mathcal{U}|)$ & $O(k*|\mathcal{U}|)$ & 3-4 hours\\  \bottomrule
\end{tabular}
\vspace{-13pt}
\end{table} 

\subsection{Fast Prediction}
Due to the network configuration of tower separation, the Two-Tower model has good advantages in predicting efficiency. Even with the addition of VKG for better performance, MVKE still retains the characteristic of fast prediction.

In Two-Tower model, the towers predict their own embeddings in parallel, contain all $|\mathcal{U}|$ users and all $|\mathcal{T}|$ tags. After predicting, it's much fast to calculate the scores between users and tags in parallel by vector dot product, even for all $|\mathcal{U}|*|\mathcal{T}|$ pairs. Finally, several top tags are selected for users according to the prediction scores. What's more, some fast vector retrieval techniques, such as FAISS~\cite{johnson2019billion}, can also be directly used for tag selection with higher efficiency. Therefore, the biggest bottleneck of prediction is the time complexity of neural network inference, and Two-Tower reduces $O(|\mathcal{U}|*|\mathcal{T}|)$ inference to $O(|\mathcal{U}|+|\mathcal{T}|)$, nearly $O(|\mathcal{U}|)$, for $|\mathcal{U}|>>|\mathcal{T}|$. And the space complexity is also near $O(|\mathcal{U}|)$.

As for MVKE, the final user embedding depends on tag embedding, so the users' embedding is not certain. However, the VKE outputs are certain, and they can be combined into the final results, with the attention weights between VKs and Tag embeddings. Therefore, in MVKE prediction phase, all of VKE outputs in user tower are stored. As for tag tower, the tag embeddings of all tasks and the tag-specific attention weights in VKGs are stored. When calculating the scores, the attention weights are first used with VKE outputs to get the final tag-specific user embedding. The next steps are the same as mentioned above. Therefore, the inference time complexity is also $O(|\mathcal{U}|)$, but its space complexity increases to $O(k*|\mathcal{U}|)$, where $k$ is the number of VKEs.

In sum, the time complexity, space complexity, and actual time cost are all shown in Table~\ref{tab-complexity}. The shown actual time is base on the computing power of four GPUs, about 80,000 queries per second (QPS), in our real business scenarios. From the table, MVKE maintains excellent time complexity, and gets good performance but sacrifices space complexity relatively. For storage resources, in fact, it is relatively sufficient, so the k times of storage cost growth is acceptable. Note that the above discussion is based on a single task. MVKE on N tasks can even save another $(N-1)$-times compute resources comparing with the basic solution, because in which each task needs its own model.

\subsection{Analysis and Discussion}
The introduction of Virtual Kernel (VK) is the core optimization point of MVKE. In the following, we will discuss the unique advantages VK brings by comparison to other models.

Firstly, compared to the single-structured tower in the basic model, VK can help to represent the user more accurately. The users' preferences are multi-faceted, varying on different topics and actions, so every facet of preferences should be represented by an individual embedding, which is a more realistic assumption. However, the basic model only outputs one compact embedding for all topics, limiting the model's modeling capabilities. MVKE can make up for these deficiencies well by assigning all the preferences with a specified number of implicit VKs, which are used to guide user features aggregation.

Secondly, the experts in classic MMoE are only structurally separated, whose ability to differentiate is not explicitly defined. However, different expert networks guided by VKs in MVKE can be more differentiated and focus on modeling one specific facet of the user's preference.

Thirdly, one of the shortcomings of Two-Tower model is that the feature interaction between the two towers is insufficient. VKs in MVKE act as a bridge to strengthen information fusion between towers, as VKs interact with both towers. While improving the performance of tasks, it can also maintain high efficiency. Therefore, the design of VK may help Two-Tower model have a broader application insight.

\section{Experiments}
In this section, we will evaluate MVKE on offline and real business scenario datasets, and compares it with several widely-used baseline methods.\footnote{We release the source code at: \url{https://github.com/MVKE2021/MVKE}, and open more detailed experimental results in the supplementary file due to limited space here.}
We will start with details about the experimental setup, including data descriptions, compared models, and specific experiment settings. 
After that, we will analyze the performance of MVKE in both offline and online scenarios to demonstrate its effectiveness and high efficiency.

\subsection{Experimental Setup}

\noindent\textbf{Dataset:} During our survey, no public dataset with multiple action labels for user tagging is found. To evaluate MVKE model, we collect users' action records from online Tencent Ads Logs to build datasets and sample a subset from all the extracted data as offline evaluation benchmarks. For the sake of privacy security, we are temporarily unable to release this data but may open it in the future. 

Specifically, we build the dataset follow below steps: 
1) Collect the user's action (click and conversion) records from raw logs, stored as \emph{"user, ads, action"} triplets. 
2) Randomly sample some triplets as negative samples, then join the records and transform them to \emph{"user, ads, click\_label, conversion\_label"}. 
3) Join user features and tags of ads from feature warehouse, stored as \emph{"user\_features, ads\_tags, click\_label, conversion\_label"}.
When model training, user features are fed into User-Tower while the tags feature are fed into Tag-Tower, and the two labels are used for loss function computation.
The online data collection is streaming, and the dataset building is updated daily. 
When evaluating the performance offline, we collect the data on two adjacent dates, and data on the former date is regarded as training data when the latter is test data. Bases on it, we get two offline datasets in different sizes by sampling. When evaluating MVKE by online A/B Test, both the model and user tag profile are updated daily.

\begin{table}
\vspace{-10pt}
\caption{Statistics of datasets. }
\vspace{-10pt}
\label{tab-data}
\centering
\small
\begin{tabular}{l|ccc}\toprule
\textbf{Statistic} & \textbf{Offline-small} & \textbf{Offline-large} & \textbf{Online}\\ \hline
\#Tag & \multicolumn{3}{c}{1,780}\\
\#Field & \multicolumn{3}{c}{User: 26, Tag: 1} \\
\#User & 100M & 500M & 1,000M\\
\#Impression & 18M & 90M & 3,600M \\
\#Click & 7M & 35M & 1,400M \\
\#Conversion & 2.2M & 11M & 440M \\ \bottomrule
\end{tabular}
\vspace{-12pt}
\end{table} 

Table~\ref{tab-data} lists some detailed data statistics. There are 26 feature fields for user tower, containing basic attributes like \emph{age, gender, education}, historical behaviors like \emph{shopping, reading}, and historical statistical features like \emph{average ctr, average cvr} etc. And there is only one field designed for tag tower, input with categorical tag IDs.

\noindent\textbf{Competitors:} For offline evaluation, we compare MVKE model with the following baseline models:
\begin{itemize}
    \item \emph{noMTL}, which is the aforementioned basic solution without Multi-Task applying. In this setting, each task employs one Two-Tower model to mine user's interest or intention tags. It is also what we run in the online system before.
    \item \emph{Hard Sharing} \cite{caruana1997multitask}, which is a straightforward way to construct an MTL model, whose bottom layer parameters for each task are shared and upper layer parameters are unique.
    \item \emph{MoE} \cite{jacobs1991adaptive}, the first gate-based selection model where all tasks share one gate, which is used to combine the experts selectively.
    \item \emph{MMoE} \cite{ma2018modeling}, a recent popular model with multiple gates to select experts, that each of them serves one task.
    \item \emph{CGC} \cite{tang2020progressive}, which is the single-level version of PLE. Note that the models in experiments are all single-level, and MVKE can also be designed to multi-level version, which belongs to our future work. 
\end{itemize}
For ablation study, we set two MVKE models for comparison:
\begin{itemize}
    \item \emph{MVKE-st}, the version of MVKE on the single tagging task. The model is evaluated twice, respectively on interest tagging and intention tagging task.
    \item \emph{MVKE-mt}, run on multiple tagging tasks. The model is applied to learn interest and intention tagging task unitedly.
\end{itemize} 
As the Tag Tower input is relatively simple, the MTL is only applied in User Tower, which is usually in combination with DeepFM to process complicated user feature input. And the Tag Tower structures are the same among all the competitors, using the hard sharing strategy. 

When evaluating MVKE on the online advertising system, the baseline method is noMTL. We conduct an online A/B Test to evaluate the effectiveness of MVKE: A specified proportion of users are bound with MVKE-produced profiles, and an equal proportion of users are bound with baseline profiles. 

\noindent\textbf{Metrics:} The metrics we focus on in the offline and online evaluation phases are different. In the offline phase, we adopt Area under the ROC curve (AUC) to evaluate the performance on two tasks, containing pCTR and pCVR. In the online phase, Tencent Ads Experimental System supplies many business metrics to reflect the actual online revenue. The primary metrics are listed as follows:
\begin{itemize}
    \item \emph{GMV}, Gross Merchandise Volume, which is the sum of all the product orders on the advertisers' websites. This metric shows the income of advertisers, and its enhancement means the advertisers are benefited.
    \item \emph{Adjust Cost}. The cost is the money advertiser paid to advertising platform. The Adjust Cost is adjusted based on actual cost and can be regarded as the income brought by the experimental strategy. Its enhancement shows the benefit that strategy brought to the platform.
\end{itemize}

\subsection{Offline Result}
We evaluate the offline performance of the proposed MVKE model in this subsection. To simulate the real business scenarios, we construct a training dataset by collecting the action records on one day, and a test dataset by collecting part of the records on the day after. 

\begin{figure}[!htbp]
\vspace{-5pt}
    \includegraphics[width=0.48\textwidth]{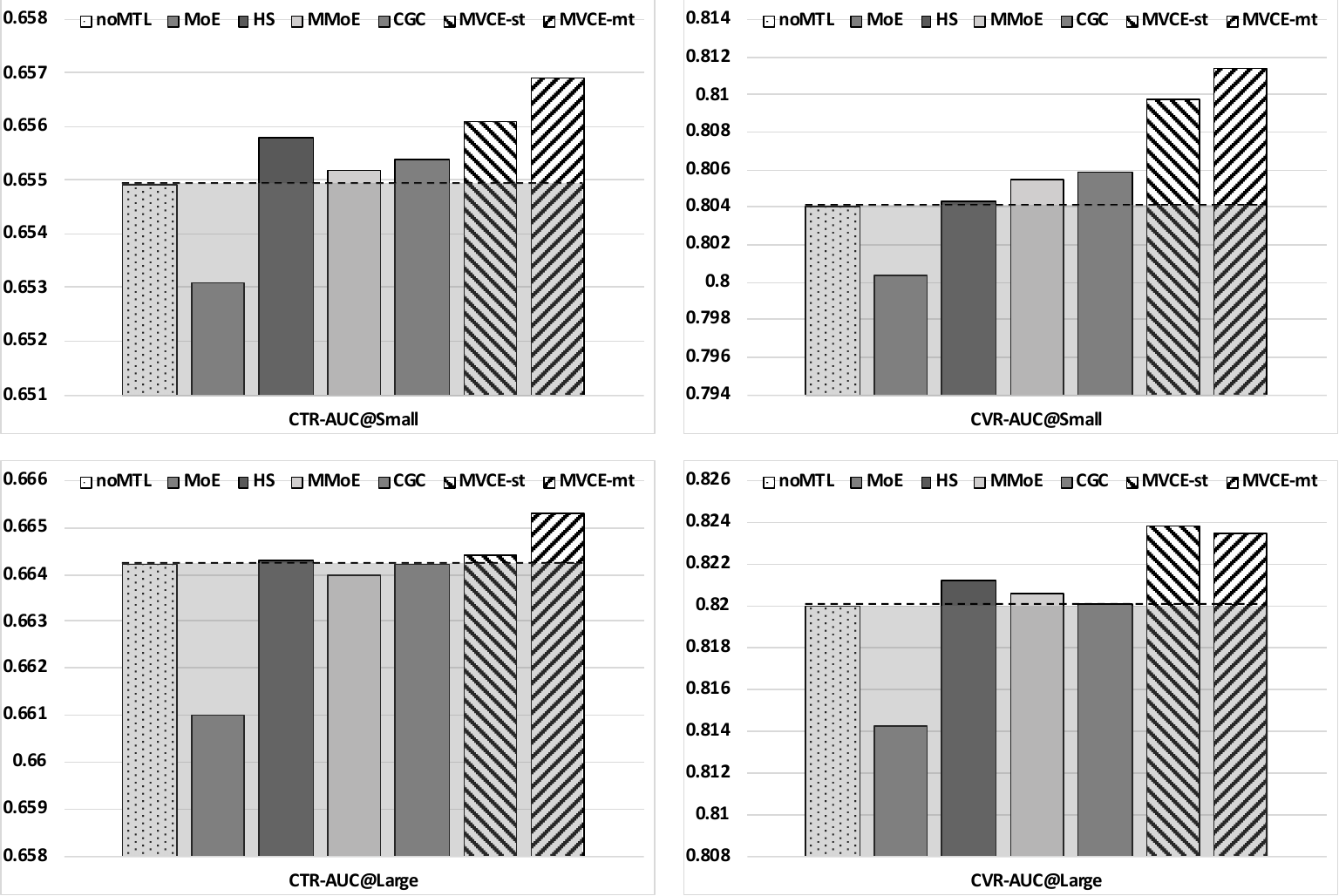}
  \vspace{-12pt}
  \caption{Offline performance comparison. AUC (higher is better) is used as metric.} 
  \label{fig:offline}
  \vspace{-10pt}
\end{figure}

The overall comparison results could be found in Fig~\ref{fig:offline}. Note that two independent models are needed to construct in no-MTL and MVKE-st, on interest tagging and intention tagging, respectively. Thus their resource consumption is also doubled compared to MTL methods. From the figure, we have the following observations:
\begin{itemize}
    \item Among all the methods, MVKE models get the best two performances on both datasets. Except for MoE model, the other MTL models outperform the model run on a single task in most cases. MoE always performs below noMTL limited by its single gate design~\cite{ma2018modeling}, no exception here.
    \item While MVKE-st is applied on the single tagging tasks, it can not only outperform noMTL about 0.1\% to 0.7\%, but also even outperform all of the MTL models much. It shows MVKE could handle the modeling on topic dimension (mentioned \textbf{C2} challenge) well and benefit much from it.
    \item In general, MVKE-mt outperforms noMTL about 0.2\% to 0.9\% and is better than all the other MTL models. Meanwhile, the others are not always better than their own competitors. So it can be concluded that MVKE is as good at modeling on action dimension (mentioned \textbf{C3} challenge).
    \item As for the comparison of MVKE-mt and MVKE-st, MVKE-mt defeats MVKE-st in most cases. The only exception is  CVR-AUC of MVKE-mt on offline-large dataset is a little bit worse than the one of MVKE-st. Thus, we can draw a conclusion that MVKE-mt can get comparable or better performance compared to MVKE-st as expected.
\end{itemize}





\subsection{Online A/B Test Result}
As an industrial model, MVKE model is fully evaluated on online experiments for about one month. In Tencent Ads, we conduct an A/B test for evaluating the effect of user profile tags, which means a certain proportion of the users are sampled to use new profiles while an equal proportion of ones use old profiles. Under the strict online experiments requirements, we successfully continue to push the proportion of using the new profiles from  1\% to 100\%, and the online performance of 1\%, 10\%, and 50\% phases are recorded.

In general, there are about 4 strategy candidates: 1) \emph{Control}, where the tags used are produced by the basic solution introduced in Section~\ref{sec:basic-solution}, and it is what already run on the system. 2) \emph{MVKE-interest}, to only test the interest tags mined by MVKE model. 3) \emph{MVKE-intention}, to only test the intention ones. 4) \emph{MVKE-both}, to test both the interest tags and intention tags mined by MVKE.

\begin{table}[!htbp]
    \centering
\small
\vspace{-5pt}
    \setlength\tabcolsep{3.5pt}
    \caption{Online performances of A/B test. The metrics lift compared with the control strategy is reported (higher is better), and the top results in different proportion groups are marked \textbf{bold}. All of the metrics reported here have passed the statistical significance test after running enough days in the experimental system.}
    \vspace{-10pt}
    \label{tab:online}
    \begin{tabular}{c|c|ccc}
    \toprule
    \textbf{Proportion} & \textbf{Treatment} & \textbf{GMV Lift}  & \textbf{Adjust Cost Lift} \\
    \hline
    \multirow{3}*{1\%} & MVKE-interest & +0.11\% & +0.14\% \\
    ~ & MVKE-intention & +0.33\% & +0.19\% \\
    ~ & MVKE-both & \textbf{+0.48\%} & \textbf{+0.55}\% \\
    \hline
    \multirow{2}*{10\%} & MVKE-intention & +0.49\% & +0.24\% \\ 
    ~ & MVKE-both & \textbf{+0.92\%} & \textbf{+0.38\%} \\
    \hline
    50\% & MVKE-both & \textbf{+0.79\%} & \textbf{+0.51\%} \\ 
    \bottomrule
    \end{tabular}
    \vspace{-10pt}
\end{table}

Under the strict requirement of the Tencent online experimental system, the bigger the proportion is set, the fewer treatment strategies can be tested. The overall results could be found in Table~\ref{tab:online}, and we have the following observations:
\begin{itemize}
    \item In general, all of the user profile tags produced by MVKE outperform online ones. Both interest tags and intention tags bring positive effects, and the strategy using both of them gets the best results. 
    \item In the 50\% phase, MVKE helps GMV lifts 0.79\%, and Adjust Cost lifts 0.51\%, which means both the advertisers and ads platform increase revenue by millions of RMB every day.
    \item As an observation on ablation study, the lift of MVKE-both is better than the one of MVKE-intention, while MVKE-intention is better than the one of MVKE-interest. Therefore, MVKE benefits intention tagging (pCVR) task more. We guess the pCVR task gets more benefits from MTL because its data sparsity is improved.
\end{itemize}

\section{Conclusion}
This paper proposes an MVKE model for user tagging tasks, building a more accurate and diverse user profile, and finally improving Tencent advertising performance much more. There are mainly three big challenges in real scenarios: the large scale of data, various topic objectives, and action objectives. Inspired by the idea of the Two-Tower model and MTL model, we propose constructing a multi-task model with high efficiency in industry. MVKE is a novel proposed model that can handle two-dimensional modeling objectives with good effectiveness and efficiency. Virtual-Kernel Experts (VKE) are introduced in MVKE model, each of which focuses on modeling one face of user preferences. The Virtual-Kernel Gate (VKG) is to combine VKE outputs selectively, under the guidance of tag tower output. Powered by these two key components, the accuracy and diversity of users' preferences are enhanced.
\bibliographystyle{ACM-Reference-Format}
\bibliography{mvke-ref}

\appendix

\begin{table*}[!htbp]
    \centering
    \small
    \caption{Taxonomy examples of level-1 caegory "Real estate".}
 \vspace{-10pt}
    \begin{tabular}{c|c|c|c}
    \toprule
    Category ID & Name & Father Node & Level  \\
    \midrule
    5 & Real estate & 0 & 1 \\
    \hline
    500 & Real estate - other properties & 5 & 2 \\
    \hline
    504 & Real estate -- real estate transaction & 5 & 2 \\
    50400 & Real estate -- real estate transaction -- real estate transaction others & 504 & 3 \\
    50401 & Real estate -- real estate transaction -- ordinary residential transaction & 504 & 3 \\
    50402 & Real estate -- real estate transaction -- villa luxury transaction & 504 & 3 \\
    50403 & Real estate -- real estate transaction -- commercial housing transaction & 504 & 3 \\
    5040300 & Real estate -- real estate transaction -- commercial housing transaction -- others & 50403 & 4 \\ 
    5040301 & Real estate -- real estate transaction -- commercial housing transaction -- office building & 50403 & 4 \\
    5040302 & Real estate -- real estate transaction -- commercial housing transaction -- bottom merchant & 50403 & 4 \\
    \hline
    505 & Real estate -- house lease & 5 & 2 \\
    50500 & Real estate -- house lease -- others & 505 & 3 \\
    50501 & Real estate -- house leasing -- ordinary residential leasing & 505 & 3 \\
    5050100 & Real estate -- house leasing -- ordinary residential leasing -- others & 50501 & 4 \\
    5050101 & Real estate -- house lease -- ordinary residential lease -- whole lease & 50501 & 4 \\
    5050102 & Real estate -- house lease -- ordinary residential lease -- joint lease & 50501 & 4 \\
    5050103 & Real estate -- house lease -- ordinary residential lease -- monthly rent & 50501 & 4 \\
    50502 & Real estate -- house lease -- villa luxury house lease & 505 & 3 \\
    50503 & Real estate -- house leasing -- commercial housing leasing & 505 & 3 \\
    5050300 & Real estate -- house leasing -- commercial housing leasing -- others & 50503 & 4 \\
    5050301 & Real estate -- house leasing -- commercial housing leasing -- shared office & 50503 & 4 \\
    \hline
    506 & Real estate - real estate developers & 5 & 2 \\
    \hline
    507 & Real estate - online real estate platform & 5 & 2\\
    \bottomrule
    \end{tabular}
    \label{tab:taxonomy}
\end{table*}
\section{Reproducibility Details}
We released the source code at: \url{https:// github.com/MVKE2021/MVKE}, where core implementation codes exist. Here, we use supplementary material to provide some important details about some specific settings, and some intuitive results in an example figure.

Generally, in our experiments, the implementation is based on Tensorflow, and all the experiments are running on NVIDIA Tesla V100, whose memory is 32 GB. In prediction phase, we compute the scores between users and tags in Spark platform, which support efficient distributed computing on CPUs.

\subsection{Model implementation details}
We compare the performance of MVKE with several widely-used and state-of-the-art methods in industry:
\begin{itemize}
    \item \emph{noMTL}, which is the aforementioned basic solution without Multi-Task applying. In this setting, each task employs one Two-Tower model to mine user's interest or intention tags. It is also what we run in the online system before.
    \item \emph{Hard Sharing} \cite{caruana1997multitask}, which is a straightforward way to construct an MTL model, whose bottom layer parameters for each task are shared and upper layer parameters are unique. In detail, the feature embeddings are shared between two tasks but the others are not, serve for CTR \& CVR prediction individually.
    \item \emph{MoE} \cite{jacobs1991adaptive}, the first gate-based selection model where all tasks share one gate, which is used to combine the experts selectively. There are 5 experts set in experiments, sharing one gate to aggregation.
    \item \emph{MMoE} \cite{ma2018modeling}, a recent popular model with multiple gates to select experts, that each of them serves one task. Similar to the setting of MoE, but each task is equipped its own gate.
    \item \emph{CGC} \cite{tang2020progressive}, which is the single-level version of PLE. Note that the models in experiments are all single-level, and MVKE can also be designed to multi-level version, which belongs to our future work. In experiments, the number of experts is also 5, and the first three serve for CTR and the last four serve for CVR, which is the same as MVKE setting.
\end{itemize}
For ablation study, we set two MVKE models for comparison:
\begin{itemize}
    \item \emph{MVKE-st}, the version of MVKE on the single tagging task. The model is evaluated twice, respectively on interest tagging and intention tagging task. There set 3 VKEs for CTR and 4 VKEs for CVR in two independent models, which are trained individually.
    \item \emph{MVKE-mt}, run on multiple tagging tasks. The model is applied to learn interest tagging and intention tagging task unitedly. There set 3 VKEs for CTR and 4 VKEs for CVR, but in total 5 VKEs, which means 2 shared VKEs are inside.
\end{itemize} 
As the Tag Tower input is relatively simple, the MTL is only applied in User Tower, which is usually in combination with DeepFM to process complicated user feature input. And the Tag Tower structures are the same among all the competitors, using the hard sharing strategy. 

We take Adam optimizer\cite{kingma2014adam} to optimize the learning, and set batch size to be 2048. For offline model for evaluation, we train it on the dataset directly. For online model for A/B test, it is firstly trained on one month data fully, and then trained incrementally on every day data. More training details of hyperparameter settings are listed in Table~\ref{tab:detail}. 

\begin{table}[!hbtp]
    \centering
    \small
    \caption{Pre-training hyperparameter settings.}
 \vspace{-10pt}
    \begin{tabular}{lc}
       \toprule
       \textbf{Hyperparameter} & \textbf{Training Value} \\
       \hline
       Learning rate & 3e-5 \\
       Adam $(\beta1, \beta2)$ & (0.9, 0.999) \\
       Batch size & 2048 \\
       Offline epoch & 2 \\
       Online epoch & 1 \\
       Feature embedding dimension & 30 \\
       Feature intersection & DeepFM \\
       DeepFM units & [512,256,128] \\
       Output embedding dimension & 30 \\
       Top layer units & [256,128,30] \\
       \bottomrule
    \end{tabular}
    \label{tab:detail}
\end{table}

\subsection{Dataset details}

\noindent\textbf{User Features} The statistics of dataset is listed in the paper, and here we introduce more detailed features used in dataset. 1) Basic attributes, including age, gender, marriage status, education, profession, consumption ability, working status, living status, etc. 2) Statistics, including average of historical CTR/CVR. 3) Behaviors, including reading, purchasing categories or keywords in more scenes like ecommerce, news, search or social scenes. During the feature extraction, we bucket the features and map the results hashly, to enhanced the performance. 

\noindent\textbf{Tag Taxonomy} There is a well-defined taxonomy for tags in our scene, contains 1786 tags in total. There are four levels in the taxonomy, where contains 18 level-1 tags, 216 level-2 tags, 942 level-3 tags and 610 level-4 tags. Taking level-one tag "Real estate" as example, there define a series of sub-categories under it, whose details are listed in Table~\ref{tab:taxonomy}.

\section{Hyperparamter sensitivity study}

\begin{figure}[!htbp]
    \includegraphics[width=0.48\textwidth]{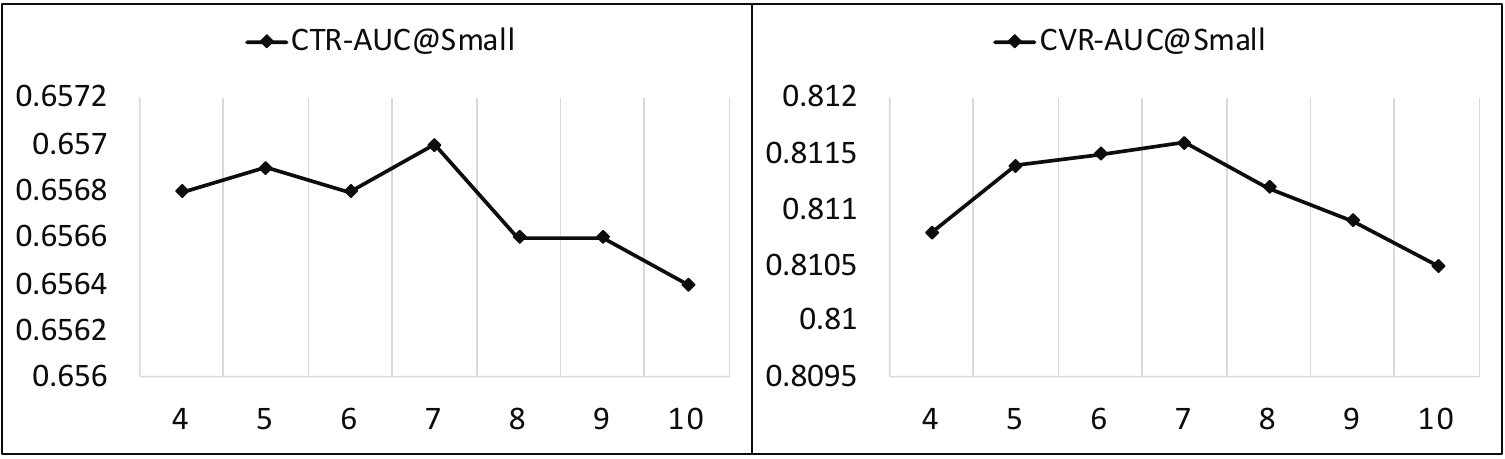}
  \caption{AUC of MVKE on different VKE number settings.} 
  \label{fig:sensitive}
\end{figure}

As shown in Fig.~\ref{fig:sensitive}, we also test the performance varying on the number of VKE for sensitivity study. For the VKE number increases from 4 to 10 in MVKE-mt on the offline-small dataset, all the models outperform the baseline models. The AUC goes higher first and then lower, and its peak is about 7, where their results fluctuate within 0.1\%. 

\subsection{Attention visualization}

\begin{figure}[!htbp]
    \includegraphics[width=0.48\textwidth]{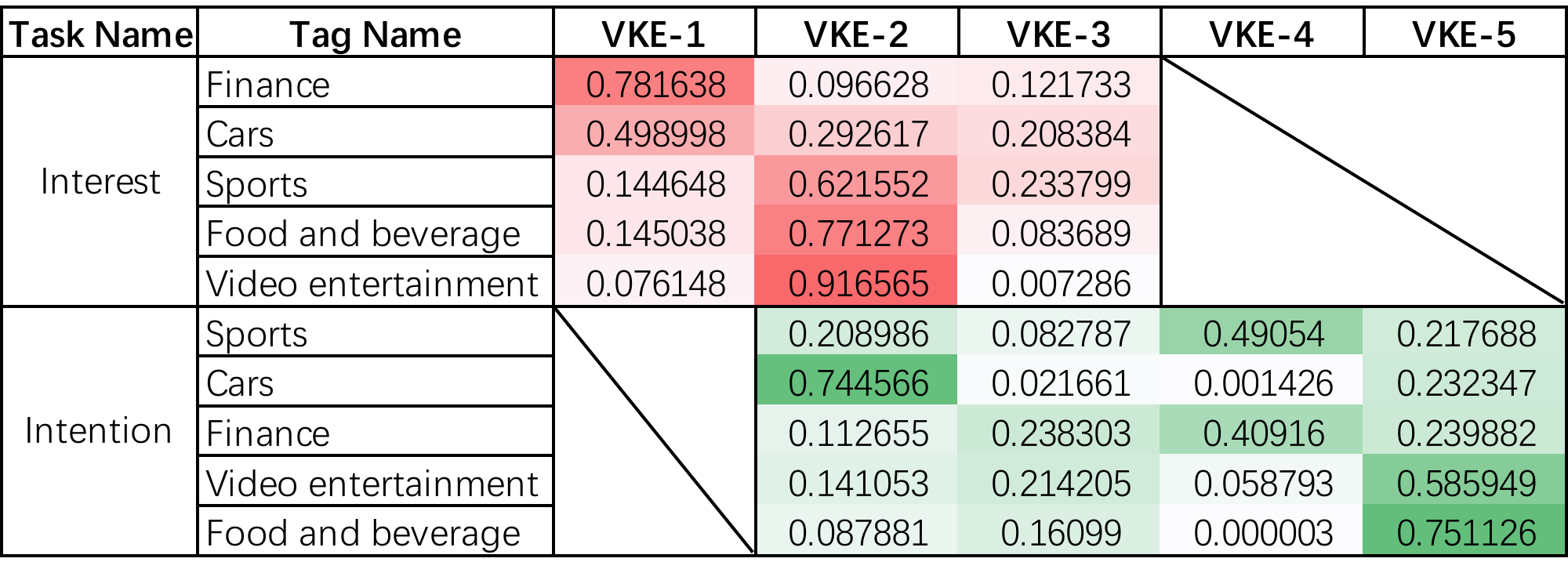}
  \caption{The weights visualization for VKE outputs combination in VKG. The number of VKE is 5 in total, the first 3 for interest tagging and the last 4 for intention tagging. The larger the weight is, the deeper the color is.} 
  \label{fig:atten-vis}
\end{figure}

Besides, we visualize the attention weights for the combination of VKE outputs after training, which is shown in Fig.~\ref{fig:atten-vis}. From the distribution, it's easy to find the weights of VKEs vary on different tags, which is as expected. 

\subsection{Detailed online results}

\begin{table}[!htbp]
    \centering
    \caption{The online performance on C-Sets, the treatment is MVKE-both.}
    \begin{tabular}{c|c|c}
    \toprule
    C-Set & GMV Lift & Adjust Cost Lift \\
    \hline
        Interest Tagging & 4.68\% & 1.96\% \\
        Intention Tagging & 24\% & 13.41\% \\
        Whole Tagging & 4.19\% & 1.82\%\\
    \bottomrule
    \end{tabular}
    \label{tab:cset}
\end{table}
For further introduce the effectiveness of MVKE on the online advertising platform, Table~\ref{tab:cset} lists the metrics on C-Set when the porpotion is 10\%. C-Set means the exact affected revenue set, where users tagging affects the revenue of platform. C-Set is a part of whole market, its metrics can more obviously reflect the effectiveness of the new treatment. It shows that the intention tagging revenue is improved much more, and all of the sets are improved obviously.

\end{document}